\documentclass[a4paper]{report}
\usepackage[utf8]{inputenc}
\usepackage[T1]{fontenc}
\usepackage{RJournal}
\usepackage{amsmath,amssymb,array}
\usepackage{booktabs}

\newlength{\cslhangindent}
\newlength{\csllabelwidth}
\newlength{\cslentryspacingunit} 
  {}%
  {\par}
\newenvironment{CSLReferences}[2] 
 {
  \setlength{\parindent}{0pt}
  \ifodd #1
  \let\oldpar\par
  \def\par{\hangindent=\cslhangindent\oldpar}
  \fi
  \setlength{\parskip}{#2\cslentryspacingunit}
 }%
 {}
\usepackage{calc}

\usepackage{subfig} \usepackage{tabu} \usepackage{longtable} \usepackage{colortbl} \usepackage{xcolor}

\begin{document}

\sectionhead{Contributed research article}
\volume{XX}
\volnumber{YY}
\year{20ZZ}
\month{AAAA}

\begin{article}
\title{vivid: An R package for Variable Importance and Variable Interactions Displays for Machine Learning Models}
\author{by Alan Inglis, Andrew Parnell, and Catherine Hurley}

\maketitle

\abstract{%
We present \CRANpkg{vivid}, an R package for visualizing variable importance and variable interactions in machine learning models. The package provides heatmap and graph-based displays for viewing variable importance and interaction jointly and partial dependence plots in both a matrix layout and an alternative layout emphasizing important variable subsets. With the intention of increasing a machine learning models' interpretability and making the work applicable to a wider readership, we discuss the design choices behind our implementation by focusing on the package structure and providing an in-depth look at the package functions and key features. We also provide a practical illustration of the software in use on a data set.
}

\hypertarget{introduction}{%
\section{Introduction}\label{introduction}}

\begin{table}

\caption{\label{tab:unnamed-chunk-3}Summary of a selection of R packages that can be used to assess the variable importance, variable interactions, or partial dependence and if these metrics are global or local and model-specific or model-agnostic. A brief description of available visualizations for evaluating model behavior is also provided.}
\centering
\fontsize{7}{9}\selectfont
\begin{tabular}[t]{>{}l>{\raggedright\arraybackslash}p{20em}>{\raggedright\arraybackslash}p{4em}>{\raggedright\arraybackslash}p{4em}>{\raggedright\arraybackslash}p{4em}}
\toprule
Package & Description & VIVI & Measure & Method\\
\midrule
\textbf{\cellcolor{gray!6}{\CRANpkg{vip}}} & \cellcolor{gray!6}{A general framework for analyzing the behavior of ML models. Also provides PDP based importance and ability to plot Shapley values. Built with \CRANpkg{ggplot2}.} & \cellcolor{gray!6}{Both} & \cellcolor{gray!6}{Agnostic} & \cellcolor{gray!6}{Global}\\
\textbf{\CRANpkg{iml}} & A general framework for analyzing the behavior of ML models. Ability to create lollipop, dot, and barplots. Also includes univariate and bivariate PDPs, ICE curves, LIME, and Shapley visualizations. Built with \CRANpkg{ggplot2}. & Both & Agnostic & Both\\
\textbf{\cellcolor{gray!6}{\CRANpkg{flashlight}}} & \cellcolor{gray!6}{A general framework for analyzing the behavior of ML models. Ability to plot VIVI measures using barplots. Includes univariate and bivariate PDPs, ICE curves, Global surrogate, and SHAP visualizations. Built with \CRANpkg{ggplot2}.} & \cellcolor{gray!6}{Both} & \cellcolor{gray!6}{Agnostic} & \cellcolor{gray!6}{Global}\\
\textbf{\CRANpkg{DALEX}} & A general framework for analyzing the behavior of ML models. Contains a suite of visualizations including Ceteris Paribus, Shapley, PDPs, model performance, and diagnostic plots. Built with \CRANpkg{ggplot2}. & Both & Agnostic & Both\\
\textbf{\cellcolor{gray!6}{\CRANpkg{lime}}} & \cellcolor{gray!6}{A general framework for fitting a local interpretable model. Ability to create VImp and model visualizations using barplots and heatmaps. Can also create interactive plots. Built with \CRANpkg{ggplot2}.} & \cellcolor{gray!6}{VImp} & \cellcolor{gray!6}{Agnostic} & \cellcolor{gray!6}{Local}\\
\addlinespace
\textbf{\CRANpkg{pdp}} & A general framework for constructing PDPs from various types machine  learning models, bivariate, and  trivariate PDPs and ICE curves.  Built with \CRANpkg{ggplot2}. & VInt & Agnostic & Global\\
\textbf{\cellcolor{gray!6}{\CRANpkg{ICEbox}}} & \cellcolor{gray!6}{Used to create Individual Conditional Expectation (ICE) plots.  Provides univariate and bivariate  PDPs and ICE curves.  Built with \CRANpkg{ggplot2}.} & \cellcolor{gray!6}{VInt} & \cellcolor{gray!6}{Agnostic} & \cellcolor{gray!6}{Local}\\
\textbf{\CRANpkg{randomForestExplainer}} & Contains a set of model-specific tools to determine which random forests variables are most important. Ability to create VIVI plots displaying the mean minimal depth distribution and conditional minimal depth. Can also display multi-way importance, pairs plots containing different metrics, and Bivariate PDP. Built with \CRANpkg{ggplot2}. & Both & Specific & Global\\
\textbf{\cellcolor{gray!6}{\CRANpkg{randomForest}}} & \cellcolor{gray!6}{Used to build random forest models. Offers VImp, error rate, and univariate PDPs. Built using base R.} & \cellcolor{gray!6}{VImp} & \cellcolor{gray!6}{Specific} & \cellcolor{gray!6}{Global}\\
\textbf{\CRANpkg{EIX}} & Contains a set of model-specific tools to determine which GBM variables are most important. Ability to create VIVI plots using lollipops, barplots, and heatmaps. Can also display dot and radar plots. Built with \CRANpkg{ggplot2}. & Both & Specific & Global\\
\addlinespace
\textbf{\cellcolor{gray!6}{\CRANpkg{varImp}}} & \cellcolor{gray!6}{Computes random forest VImps for the conditional inference random forest of the \CRANpkg{party} package.} & \cellcolor{gray!6}{Vimp} & \cellcolor{gray!6}{Specific} & \cellcolor{gray!6}{Global}\\
\textbf{\CRANpkg{bartMachine}} & Used to build Bayesian additive regression tree models. Ability to plot VIVI measures with uncertainty included using barplots. Also includes a suite of model diagnostic  plots and univariate PDP. Built using  base R. & Both & Specific & Global\\
\bottomrule
\end{tabular}
\end{table}

Our motivation behind the creation of the \CRANpkg{vivid} package is to investigate machine learning
models in a way that is simple to understand while also offering helpful
insights into how variables affect the fit. We do this through the use
of heatmaps, network graphs, and both a generalized pairs plot style
partial dependence plot (PDP) (Friedman 2000) and a space saving PDP based
on key variable subsets. While the techniques and fundamental goals of
these visualizations have been discussed in Inglis, Parnell, and Hurley (2022a), we
focus here on the implementation details of the package by providing a
complete listing of the functions and arguments included in the \CRANpkg{vivid}
package with further examples indicating advanced usage beyond that
previously shown. In this work we examine the decisions made when
designing the package and provide an in-depth look at the package
functions and features with the intention of making the work applicable
to a larger readership. This article outlines the general architectural
principles implemented in \CRANpkg{vivid}, such as the data structures we use and data
formatting, function design, filtering techniques, and more. We
illustrate each function by way of a practical example. Our package is
available on the Comprehensive R Archive Network at
\url{https://cran.r-project.org/web/packages/vivid} or on GitHub at
\url{https://github.com/AlanInglis/vivid}.

In recent years machine learning (ML) algorithms have emerged as a
valuable tool for both industry and science. However, due to the
black-box nature of many of these algorithms it can be challenging to
communicate the reasoning behind the algorithm's decision-making
processes. With the need for transparency in ML growing it is important
to gain understanding and clarity about how these algorithms are making
predictions (Antunes et al. 2018; Felzmann et al. 2019). Many R
packages are now available that aid in creating interpretable machine
learning (IML) models such as \CRANpkg{iml} (Molnar, Bischl, and Casalicchio 2018), \CRANpkg{DALEX} (Biecek 2018), and
\CRANpkg{lime} (Hvitfeldt, Pedersen, and Benesty 2022). For a
comprehensive review of IML, see Molnar (2022) and
Biecek and Burzykowski (2021).

How we choose to visualize aspects of the model output is of vital
importance in how a researcher can interpret and communicate their
findings. Consequently, model summaries such as variable importance and
variable interactions (VImp and VInt; together we term these VIVI) are
frequently used in various fields to comprehend and explain the hidden
structure in an ML fit. In ecology they are employed to determine the
causes of ecological phenomena (e.g. Murray and Conner 2009); in meteorology
VImp measures and partial dependence plots are used to examine air
quality (e.g. Grange et al. 2018); in bioinformatics, understanding
gene-environment interactions have made these measures an important tool
for genomic analysis (e.g. Chen and Ishwaran 2012).

In Table 1 we summarize VIVI measures and visualizations
provided by a selection of R packages. VIVI measures are categorized as global or local, depending on whether they refer to the entire predictor space or a specific sub-region. For instance, Individual Conditional Expectation (ICE) curves (Goldstein et al. 2015) and Local Interpretable Model-agnostic Explanations (LIME) (Ribeiro, Singh, and Guestrin 2016) are considered local methods (implementations of which can be found in the \CRANpkg{ICEbox} package (Goldstein et al. 2015) and the \CRANpkg{lime} package (Hvitfeldt, Pedersen, and Benesty 2022)), whereas partial dependence and permutation importance represent global methods. Packages that incorporate local or global methods in Table 1 can be found under the column `Method'. Additionally, VIVI measures can be broken down into model specific (embedded) methods or model agnostic methods. The packages listed in Table 1 are grouped by whether they can compute model specific or agnostic measures and can be found under the column `Measure'. In embedded methods the variable importance is incorporated into the ML algorithm. For example, random forests (RF)
(Breiman 2001) and gradient boosting machines (GBM) (Friedman 2000) use the tree structure to evaluate the performance of the model. Bayesian additive regression tree models (BART) (Chipman, George, and McCulloch 2010) also use an embedded method to obtain VIVI measures by looking at the
proportion of splitting rules for variables or variable pairs used in the trees. The package \CRANpkg{randomForestExplainer} (Paluszynska, Biecek, and Jiang 2020) provides a set of tools to understand what is happening inside a random forest and uses the concept of minimal depth (Ishwaran et al. 2010) to assess both importance and interaction strength by examining the position of a variable within the trees. Additionally the \CRANpkg{varImp} (Probst 2020) package can be used to compute importance scores for the conditional inference random forest of the \CRANpkg{party} package (Strobl et al. 2008). For gradient boosted machines, the \CRANpkg{EIX} (Maksymiuk, Karbowiak, and Biecek 2021) package can be used to measure and identify VIVI and visualize the results.

Model-agnostic methods are techniques that can, in principle, be applied
to any ML algorithm. Agnostic methods not only provide flexibility in
relation to model selection but are also useful for comparing different
fitted ML models. An example of a model agnostic approach for evaluating
VImps is permutation importance (Breiman 2001). This method
calculates the difference in a model's predictive performance following
a variable's permutation; implementations are available in the packages \CRANpkg{iml}, \CRANpkg{flashlight}, \CRANpkg{DALEX} and \CRANpkg{vip} (Greenwell and Boehmke 2020). Each package provides options for specifying the performance metric to use in computing the model performance as well as providing options to select the number of permutations, with \CRANpkg{flashlight} and \CRANpkg{DALEX} packages additionally providing an option to select the number of observations that should be sampled for the calculation of variable importance. The \CRANpkg{vip} package also provides a PDP/ICE-based variable importance method, which computes quantifies the variability in PDP and ICE plots. For VInts, Friedman's \(H\)-statistic (Friedman and Popescu 2008) is an agnostic interaction measure derived from the partial dependence by comparing a pair of variables' joint effects with the sum of their marginal
effects. Packages \CRANpkg{iml} and \CRANpkg{flashlight} provide implementations.

Partial dependence plots (PDPs) were first introduced by Friedman (2000) as
a model agnostic way to visualize the relationship between a specified
predictor variable and the fit, averaging over other predictors'
effects. Similar to PDPs, individual conditional expectation curves (ICE)
(Goldstein et al. 2015) show the relationship between a specified predictor and the
fit, fixing the levels of other predictors at those of a particular
observation. PDP curves are then the average of the ICE curves over all
observations in the dataset. R packages offering PDPs include
\CRANpkg{iml}, \CRANpkg{DALEX} (which calls the \CRANpkg{ingredients} package (Biecek and Baniecki 2023)), \CRANpkg{flashlight}, and \CRANpkg{pdp} (Greenwell 2017);
the package \CRANpkg{ICEbox} (Goldstein et al. 2015) provides ICE curves and variations. Each package provides options for selecting the size of the grid for evaluating the predictions, with \CRANpkg{flashlight} and \CRANpkg{DALEX} providing options to select the number of observations to consider in calculating the partial dependence.

In \CRANpkg{vivid} we provide a suite of functions (see Table 2) for calculating and visualizing variable importance, interactions and the partial dependence. Our displays conveniently show (either model specific and agnostic) VImp and VInt jointly using heatmaps and network graphs. Through the use of seriation techniques, we group together variables with the greatest impact on the response. For our network plot, additional filtering to remove less influential variables and clustering options to group highly interacting variables are provided, thus providing a more informative picture identifying relevant features. Additionally, our implementation makes it possible to apply our methods to subsets of data, which leads to locally-based importance measures. Our generalized PDP (GPDP) displays partial dependence plots in a matrix layout combining univariate and bivariate partial dependence plots with variable scatterplots. This layout, coupled with seriation, allows for quick assessment of of how pairs of variables have an impact on the model fit. We furthermore provide a more compact version of the GDPD, the so-called zen-partial dependence plot (ZPDP) consisting only of those bivariate partial dependence plots with high VInt. All of our displays are designed to quickly identify how variables, both singly and jointly, affect the fitted response and can be used for regression or classification fits. As the output of our displays are \CRANpkg{ggplot2} objects (Wickham 2016), they are easily customizable and provide the flexibility to create custom VIVI visualizations.

\begin{table}

\caption{\label{tab:unnamed-chunk-7}Summary of functions available in the vivid package. 
The main construction function is vivi which is used to calculate the VIVI values 
for subsequent use in the visualizations.}
\centering
\fontsize{8}{10}\selectfont
\begin{tabular}[t]{>{}l>{\raggedright\arraybackslash}p{20em}>{\raggedright\arraybackslash}p{10em}}
\toprule
Function & Description & Type\\
\midrule
\cellcolor{gray!6}{\texttt{vivi}} & \cellcolor{gray!6}{Create a VIVI matrix of class  \texttt{vivid}} & \cellcolor{gray!6}{VIVI construction}\\
\texttt{viviReorder} & Reorders a square matrix so high VIVI values are pushed to the top left of the matrix & VIVI construction\\
\cellcolor{gray!6}{\texttt{viviHeatmap}} & \cellcolor{gray!6}{Heatmap plot of VIVI values} & \cellcolor{gray!6}{Visualization}\\
\texttt{viviNetwork} & Network plot of VIVI values & Visualization\\
\cellcolor{gray!6}{\texttt{pdpVars}} & \cellcolor{gray!6}{Univariate partial dependence plot with ICE curves  displayed as a grid} & \cellcolor{gray!6}{Visualization}\\
\addlinespace
\texttt{pdpPairs} & Pairs plot showing bivariate PDP, ICE/univariate PDP,  and data & Visualization\\
\cellcolor{gray!6}{\texttt{pdpZen}} & \cellcolor{gray!6}{A zigzag expanded navigation plot (zenplot) displaying partial dependence values} & \cellcolor{gray!6}{Visualization}\\
\texttt{CVpredictfun} & Predict function & Utility\\
\cellcolor{gray!6}{\texttt{zPath}} & \cellcolor{gray!6}{Constructs a zenpath for connecting and displaying pairs  to be used with pdpZen} & \cellcolor{gray!6}{Utility}\\
\texttt{as.data.frame.vivid} & Takes a matrix of class vivid and turns it into a data frame & Utility\\
\addlinespace
\cellcolor{gray!6}{\texttt{vip2vivid}} & \cellcolor{gray!6}{Takes measured importance and interactions from the vip  package and turns them into vivid matrix which can be  used for plotting} & \cellcolor{gray!6}{Utility}\\
\bottomrule
\end{tabular}
\end{table}

This paper is structured as follows. First we introduce a dataset and fit models that will be used as examples throughout this paper. Following this, we describe \CRANpkg{vivid} functionality for calculating VIVI. We then move on to visualizations and focus on the functionality provided by the two functions \texttt{viviHeatmap} and \texttt{viviNetwork} for displaying VIVI, and two functions for displaying PDPs namely, \texttt{pdpPairs} and \texttt{pdpZen}. Finally we provide some concluding discussion.

\hypertarget{datamodel}{%
\section{Example: Data and Models}\label{datamodel}}

The well-known Boston housing data (Harrison Jr and Rubinfeld 1978) from the R package \CRANpkg{MASS} (Venables and Ripley 2002) concerns prices of 506 houses and 14 predictor variables including property attributes such as number of rooms and social attributes including crime rate and pollution levels. The response is the median value of owner-occupied homes in \$1000s (medv).

We first fit a random forest (using the \CRANpkg{randomForest} package). In order to avail of all the available embedded variable importance scores, the \texttt{importance} argument must be \texttt{TRUE} when calling the \texttt{randomForest} function. This allows any of the provided importance metrics to be used in \CRANpkg{vivid}. However, in our following examples, we use an agnostic VImp measure supplied by the \CRANpkg{vivid} package, which allows us to directly compare VImp values across different fits.

\begin{verbatim}
library("randomForest") 
library("MASS")
set.seed(1701)  
data("Boston")

rf <- randomForest(medv ~., data = Boston)
\end{verbatim}

Next we fit a gradient boosted machine (using the \CRANpkg{xgboost} package). For the GBM we set the maximum number of boosting iterations, \texttt{nrounds}, to 100 as no default is provided in \CRANpkg{xgboost}.

\begin{verbatim}
library("xgboost") 
gbst <- xgboost(data = as.matrix(Boston[,1:13]), 
                label =  as.matrix(Boston[,14]),
                nrounds = 100,
                verbose = 0)
\end{verbatim}

In the following sections we will explain how aspects of the two fits can be compared with \CRANpkg{vivid} software. We will also explain aspects of our software design with reference to these fits.

\hypertarget{data}{%
\section{Calculating VIVI}\label{data}}

The first step in using \CRANpkg{vivid} is to calculate variable importance and interactions for a model fit.
The \texttt{vivi} function calculates both of these, creating a square, symmetric matrix containing variable importance on the diagonal and variable interactions on the off-diagonal. Required inputs are a fitted ML model, a data frame on which the model was trained, and the name of the response variable for the fit. The returned matrix has importance and interaction values for all variables in the supplied data frame, excluding the response. Our visualizations functions \texttt{viviHeatmap} and \texttt{viviNetwork} are designed to show the results of a \texttt{vivi} calculation, but will work equally well for any square matrix with identical row and column names. (Note, the symmetry assumption is not required for \texttt{viviHeatmap} and \texttt{viviNetwork} uses interaction values from the lower-triangular part of the matrix only.)

The code snippet below shows the creation of a \texttt{vivid} matrix for the random forest fit. For clarity, we include all of the \texttt{vivi} function arguments for the random forest fit, though only the first three are required. Other inputs will be described in the section \protect\hyperlink{sec:vivimatarg}{\texttt{vivi} function additional arguments}.

\begin{verbatim}
library("vivid")

set.seed(1701)
viviRf <- vivi(fit = rf,
               data = Boston,
               response = "medv",
               reorder = FALSE,
               normalized = FALSE,
               importanceType = "agnostic",
               gridSize = 50,
               nmax = 500,
               class = 1,
               predictFun = NULL,
               numPerm = 4)
\end{verbatim}

In the absence of any model-specific importance measure we use an agnostic permutation method described by Fisher, Rudin, and Dominici (2019) to obtain the variable importance scores. In this method a model error score (root mean square error) is calculated, then each feature is randomly permuted and the model error is re-calculated. The difference in performance is considered to be the variable importance score for that feature. By default the permutation is set to be replicated four times to account for variability. However, we provide an option to select a desired number of permutations (via the \texttt{numPerm} argument)).

The \texttt{vivi} function calculates both the importance and interactions using S3 methods. By default, the agnostic importance and interaction scores in \CRANpkg{vivid} are computed using the generalized predict function from the \CRANpkg{condvis2} package (C. Hurley, OConnell, and Domijan 2022). Consequently, \CRANpkg{vivid} can be used out-of-the-box with any model type that works with \CRANpkg{condvis2} predict (see \texttt{CVpredict} from \CRANpkg{condvis2} for a full list of compatible model types). To allow \CRANpkg{vivid} to run with other model fits, a custom predict function must be passed to the \texttt{predictFun} argument (as discussed below).

The S3 method used to obtain the importance is called \texttt{vividImportance}. \texttt{vivid} relies on the \CRANpkg{flashlight} package to calculate agnostic importance via \texttt{flashlight::light\_importance} which currently works for numeric and numeric binary responses only. For model-specific variable importance, we provide individual methods to access importance scores for some of the most popular model fitting \texttt{R} packages, namely; \CRANpkg{ranger} (Wright and Ziegler 2017), \CRANpkg{randomForest}, \CRANpkg{mlr} (Bischl et al. 2016), \CRANpkg{mlr3} (Lang et al. 2019), and \CRANpkg{parsnip} (Kuhn and Vaughan 2022) (however, more could be added via additional methods). To access any available model-specific variable importance from the aforementioned packages, the \texttt{importanceType} argument must be set to equal the selected importance metric. For example, to select the percent increase in mean square error importance score from the \CRANpkg{randomForest} package, the \texttt{importanceType} argument must be set to equal this measure as it is called within the \CRANpkg{randomForest} package, that is; \texttt{importanceType\ =\ "\%IncMSE"}. If the \texttt{importanceType} is not set or is set to equal \texttt{agnostic}, then the agnostic importance is calculated. It should be noted that when comparing different model fits, using model-specific variable importance will result in importance measures that are not directly comparable (however, comparing model specific scores could be useful when comparing the same fitting procedure evaluated using different parameters). In the case that a practitioner may wish to use our visualizations to compare different model fits, we recommend using the agnostic permutation approach supplied by \CRANpkg{vivid} to make a direct comparison of the importance measures.

For variable interactions, we use the model-agnostic Friedman's \(H\)-statistic to identify any pairwise interactions. As discussed in Inglis, Parnell, and Hurley (2022a), we recommend the unnormalized version of the \(H\)-statistic which prevents detection of spurious interactions which can occur when the bivariate partial dependence function (used in the construction of the \(H\)-statistic) is flat. In the case of a binary response classification model, we follow Hastie, Tibshirani, and Friedman (2009) and compute the \(H\)-statistic and partial dependence on the logit scale.

The \texttt{vivi} function calculates interactions using an S3 method called \texttt{vividInteraction}, which again relies on the \CRANpkg{flashlight} package to calculate Friedman's \(H\)-statistic via
\texttt{flashlight::light\_interaction}. Friedman's \(H\)-statistic is the only interaction measure currently available in \CRANpkg{vivid}, though the method of Greenwell, Boehmke, and McCarthy (2018) could also be used for this purpose. Embedded interaction measures could easily be incorporated via S3 methods in future.

\CRANpkg{flashlight} simplifies the calculation of VIVI values
as it allows a custom predict function to be supplied for the calculation of agnostic importance and the \(H\)-statistic; this flexibility means importance and the \(H\)-statistic can be calculated for any ML model. Note that as \CRANpkg{flashlight} importance and interaction functions act in a model agnostic way, they will give a VIVI of zero for variables in the dataset (except the response) that are not used by the supplied ML fit.
We supply an internal custom predict function called \texttt{CVpredictfun} to both \texttt{flashlight::light\_importance} and \texttt{flashlight::light\_interaction}. \texttt{CVpredictfun} is a wrapper around \texttt{CVpredict} from the \CRANpkg{condvis2} package, which adds an option for the classification to select (via the \texttt{class} argument to \texttt{vivi}) the class to be used for prediction and calculates predictions on the logit scale by default.
\texttt{CVpredict} accepts a broad range of fit classes thus streamlining the process of calculating VIVI.

In situations where the fit class is not handled by \texttt{CVpredict} (as is the case for the GBM model created from \CRANpkg{xgboost}), supplying a custom predict function to the \texttt{vivi} function by way of the \texttt{predictFun} argument allows the agnostic VIVI values to be calculated.
In the code snippet below, we build the \texttt{vivid} matrix for the GBM fit using a custom predict function, which must be of the form given in the code snippet.
For brevity we omit some of the optional \texttt{vivi} function arguments. By default, the agnostic variable importance is used to allow for a direct comparison of the importance measures for both model fits.

\begin{verbatim}
# predict function for GBM
pFun <- function(fit, data, ...) predict(fit, as.matrix(data[,1:13]))
\end{verbatim}

\begin{verbatim}
set.seed(1701) 
viviGBst <- vivi(fit = gbst,
                 data = Boston,
                 response = "medv",
                 reorder = FALSE,
                 normalized = FALSE,
                 predictFun = pFun)
\end{verbatim}

\hypertarget{sec:vivimatarg}{%
\subsection{\texorpdfstring{\texttt{vivi} function additional arguments}{vivi function additional arguments}}\label{sec:vivimatarg}}

The \texttt{vivi} function has 11 arguments. Some of these have been discussed above, including \texttt{fit}, \texttt{data}, \texttt{response}, \texttt{importanceType}, and \texttt{predictFun}. Here we provide a summary of the remaining arguments. First, the \texttt{normalized} argument determines if Friedman's \(H\)-statistic should be normalized or not (see Inglis, Parnell, and Hurley (2022a), for the pros and cons of each version). The arguments \texttt{gridSize} and \texttt{nmax} are used to set the size of the grid for evaluating the predictions and maximum number of data rows to consider, respectively, in the calculation of the \(H\)-statistic. Lowering the grid size can provide a significant speed boost, though at the expense of predictive accuracy. Additionally, sampling the data via \texttt{nmax} offers a speed boost. The default values for \texttt{gridSize} and \texttt{nmax} are 50 and 500, respectively.

\hypertarget{sec:speed}{%
\subsection{Speed tests}\label{sec:speed}}

\begin{figure}

{\centering \includegraphics[width=0.6\linewidth]{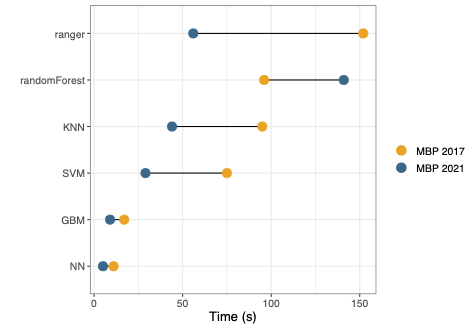} 

}

\caption{Mean time over five runs, on two MacBooks, for the creation of a vivid matrix for different models. Times are highly dependent on the model fit, with NN the fastest and random forests the slowest.}\label{fig:speedtest}
\end{figure}

A drawback of using Friedman's \(H\)-statistic as a measure of interaction is that it is a computationally expensive calculation, and may be especially time-consuming for models where prediction is slow.
Figure \ref{fig:speedtest}
shows the build time (rounded to the nearest second) averaged over five runs for the creation of a \texttt{vivid} matrix with default parameters for different ML algorithms using the Boston Housing data. As the Boston housing data has 13 predictor variables, Friedman's \(H\)-statistic is computed for 91 predictor pairs. The ML algorithms are: GBM, random forest, support vector machine (SVM), neural network (NN), and k-nearest neighbors (KNN). The SVM, NN, and KNN were built using the \CRANpkg{e1071} (Meyer et al. 2021), \CRANpkg{nnet} (Venables and Ripley 2002), and \CRANpkg{kknn} (Schliep and Hechenbichler 2016) packages, with the KNN being built through the \CRANpkg{mlr3} (Lang et al. 2019) framework. Each of the models were built using their default settings and, for each model fit, the agnostic VIVI was measured. The speed tests were performed on both a 2017 Mac-book Pro 2.3 GHz Dual-Core Intel Core i5 with 8GB of RAM and a 2021 32GB Mac-book M1 Pro. Here we are essentially comparing predict times for the various fits. The NN fit created using the \CRANpkg{nnet} package was the fastest, followed by GBM. Both random forests are the slowest, and surprisingly, the older Mac beats its higher spec cousin for the \CRANpkg{randomForest} fit.

\hypertarget{sec:vip2vivid}{%
\subsection{\texorpdfstring{Alternative construction of a \texttt{vivid} matrix}{Alternative construction of a vivid matrix}}\label{sec:vip2vivid}}

A \texttt{vivid} matrix may also be obtained from variable importance and interaction values calculated elsewhere. The package \CRANpkg{vip} offers these, and evaluates interactions using a method called the \emph{feature importance ranking measure} (FIRM, see Greenwell, Boehmke, and McCarthy (2018), for more details). The \texttt{vip2vivid} function we provide in \CRANpkg{vivid} takes VIVI values created in \CRANpkg{vip} and turns them into a \texttt{vivid} matrix, that can be subsequently used with our plotting tools. For example, in the code below, model-specific VImp and FIRM VInt scores are calculated for the random forest fit, and subsequently arranged into a \texttt{vivid} matrix with the VImps on the diagonal and VInts on the off-diagonal.

\begin{verbatim}
library("vip")
# get model specific VImps using vip package
vipVImp <- vi(rf, method = 'model')
# get VInts using vip package
vipVInt <- vint(rf, feature_names = names(Boston[-14]))

# turn into vivi-matrix
vipViviMat <- vip2vivid(importance = vipVImp, interaction = vipVInt)
\end{verbatim}

\hypertarget{sec:heatmapSec}{%
\section{Heatmap of Variable Importance and Variable Interactions}\label{sec:heatmapSec}}

The \texttt{viviHeatmap} function constructs a heatmap displaying both importance and interactions, with importance on the diagonal and interactions on the off-diagonals. A \texttt{vivid} matrix is the only required input, which does not necessarily need to be symmetric (for example, the interaction measures from the \CRANpkg{randomForestExplainer} package are asymmetric and could be visualized using our heatmap. Color palettes for the importance and interactions are optionally provided via \texttt{impPal} and \texttt{intPal} arguments.
For the default color palette we choose single-hue, color-blind friendly sequential color palettes from Zeileis et al. (2020), where low and high VIVI values are represented by low and high luminance color values respectively, aiding in highlighting values of interest.

The ordering of the heatmap is taken from the ordering of the input matrix. As \texttt{reorder} was set to \texttt{FALSE} when building both the random forest and GBM fit \texttt{vivid} matrix, the ordering of the heatmaps matches the variable order in the dataset. This is useful for directly comparing multiple heatmaps, however it does not necessarily lend itself for easy identification of the largest VIVI values. If we were to seriate both \texttt{vivi} matrices separately, we would end up with different optimal orderings for each matrix. An alternative is to create a common ordering by averaging over the two \texttt{vivid} matrix objects and applying the \texttt{vividReorder} function to the result.
(This function uses a seriation algorithm based on the techniques of Earle and Hurley (2015) designed to place high interaction variables adjacently and to pull high VIVI variables towards the top-left; see Inglis, Parnell, and Hurley (2022a) for details.)
Both VIVI matrices are then re-ordered using the newly obtained variable order.

\begin{verbatim}
# average over matrices and seriate to get common ordering
viviAvg <- (viviRf + viviGBst) / 2
viviAvgReorder <- vividReorder(viviAvg)

# reorder vivi-matrices 
ord <- colnames(viviAvgReorder)
viviRf <- viviRf[ord,ord]
viviGBst <- viviGBst[ord,ord]
\end{verbatim}

Arguments \texttt{impLims} and \texttt{intLims} specify the range of importance and interaction values to be mapped to colors. Default values are calculated from the maximum and minimum VIVI values in the \texttt{vivid} matrix. Importance and interaction values falling outside the supplied limits are squished to the closest limit. It can be useful to specify these limits in the situation where there is an extremely large VIVI value that dominates the display, or where we wish two or more plots to have the same limits for comparison purposes, as in the example below. The \texttt{angle} argument is used to rotate the x-axis labels.

\begin{figure}

{\centering \subfloat[\label{fig:heatmaps-1}]{\includegraphics[width=0.5\linewidth]{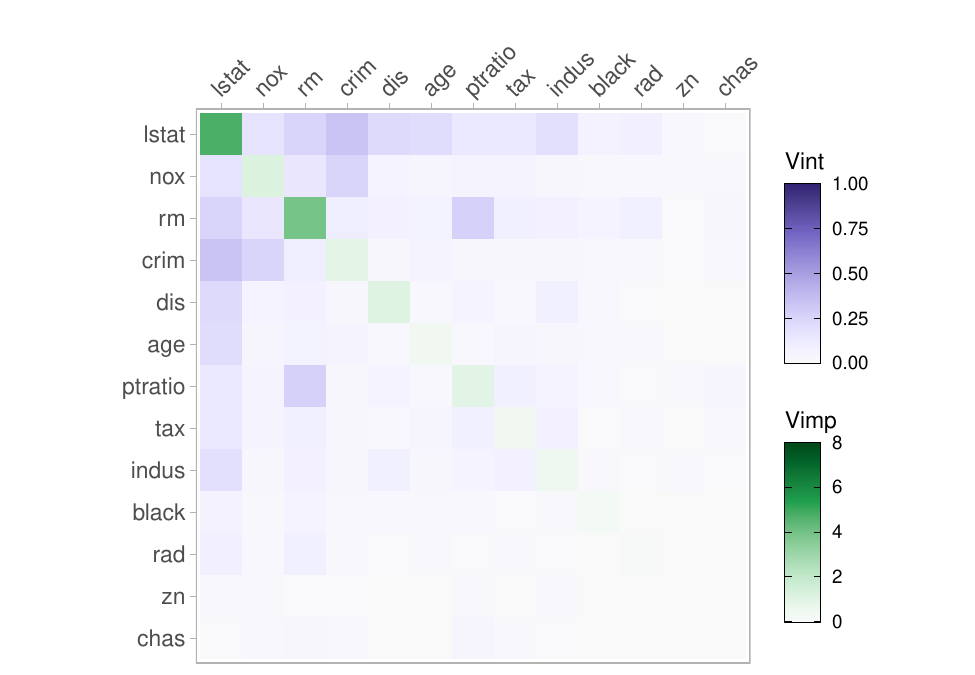} }\subfloat[\label{fig:heatmaps-2}]{\includegraphics[width=0.5\linewidth]{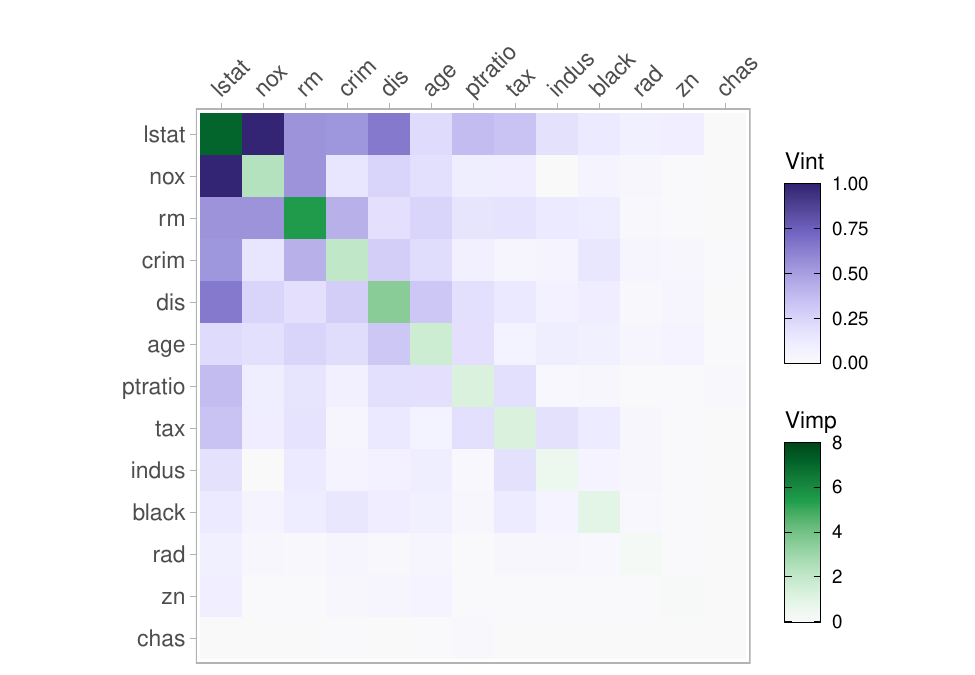} }

}

\caption{Agnostic variable importance and variable interaction scores for a random forest fit in (a) and GBM fit in (b) on the Boston housing data displayed as a heatmap. The random forest fit has weaker interactions and lower importance scores than the GBM fit. Both fits identify $lstat$ as the most important followed by $rm$. In both fits, $lstat$ has numerous interactions with other variables, notably $crim$ in the random forest fit and $nox$ in the GBM fit.}\label{fig:heatmaps}
\end{figure}

Figure \ref{fig:heatmaps} shows our improved ordering so that variables with high VIVI values are pushed to the top left of the plots. Filtering can also be applied to the input matrix to display a subset of variables. When compared to the GBM fit in (b), the random forest fit in (a) has weaker interactions and lower importance scores. Both plots identify \(lstat\) as being the most important. Both fits also show that \(lstat\) (the percentage of lower status of the population) interacts with several other variables, though the interactions are much stronger for the GBM. Notably, the strongest interaction in both fits are different. These are \(lstat:crim\) (where \(crim\) is the per capita crime rate by town) for the random forest fit and \(lstat:nox\) (where \(nox\) is parts per 10 million nitrogen oxides concentration) for the GBM fit.

\begin{verbatim}
viviHeatmap(viviRf, angle = 45, intLims = c(0,1), impLims = c(0,8))
viviHeatmap(viviGBst, angle = 45, intLims = c(0,1), impLims = c(0,8))
\end{verbatim}

\hypertarget{networkSec}{%
\section{Network of Variable Importance and Variable Interactions}\label{networkSec}}

\begin{figure}

{\centering \subfloat[\label{fig:networks1-1}]{\includegraphics[width=0.5\linewidth]{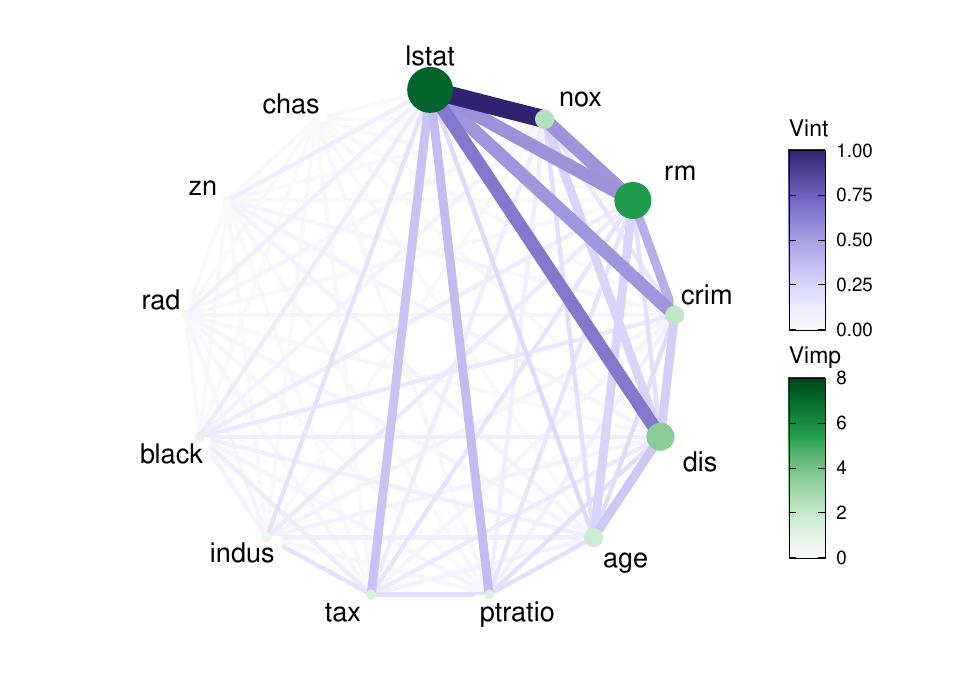} }\subfloat[\label{fig:networks1-2}]{\includegraphics[width=0.5\linewidth]{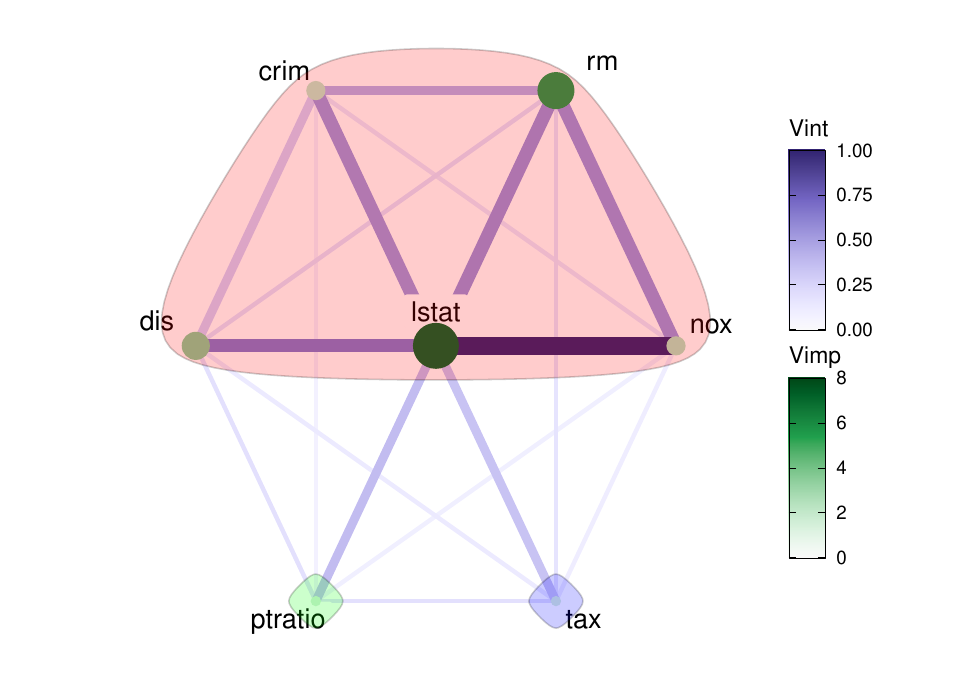} }

}

\caption{Network plots showing VIVI scores obtained from a GBM fit on the Boston housing data. In (a) we display the all values in a circle. In (b) we use a hierarchical clustering to group variable with high VIVI together and rearrange the layout using an igraph function.}\label{fig:networks1}
\end{figure}

The \texttt{viviNetwork} function constructs a network graph displaying both importance and interactions.
Similarly to \texttt{viviHeatmap} , this function takes a \texttt{vivid} matrix as the only required input and provides a visual representation of the magnitude of the importance and interaction values through the size of the nodes and edges in the graph, in addition to color. In the plot each variable is represented as a node, with its importance being represented through size and color such that larger, darker nodes indicate a higher importance. Each pairwise interaction is represented by a connecting edge, where larger interaction values get thicker, darker edges; Figure \ref{fig:networks1} provides an example. This type of plot benefits from being able to quickly identify the magnitude of the importance and interactions of the variables that have the most impact on the response. The \texttt{viviNetwork} function optional arguments follows the same conventions as \texttt{viviHeatmap}: custom color palettes for importance and interactions are provided via the \texttt{impPal} and \texttt{intPal}, and the range of VIVI values to be mapped to the colors are specified via the \texttt{impLims} and \texttt{intLims}.

By default, we choose a circular layout to display the graphs, as when coupled with the seriated \texttt{vivid} matrix, variables with high VIVI are grouped in a clock-wise arrangement starting at the top. This arrangement allows for easy identification of variables with high VIVI. Custom layouts are possible by providing a numeric matrix with two columns and one row per node to the \texttt{layout} argument. Additionally, any of the layouts available in the \CRANpkg{igraph} package (Csardi and Nepusz 2006) can be specified. The subject of network graph layouts has been extensively studied (for examples see Purchase (1997), Herman, Melançon, and Marshall (2000), Freeman (2000)). It has been shown that certain layouts of network graphs can significantly aid in cognitive interpretation. For example, Purchase (1997) note that reducing the number of edge crossovers is by far the most important aesthetic (even for small amounts of data), while maximizing symmetry has a lesser effect. Several of the layouts provided by the \CRANpkg{igraph} package can aid interpretation (such as the Sugiyama layout algorithm (Sugiyama, Tagawa, and Toda 1981), which tries to minimize edge crossover.)

We provide options to filter the graph via the \texttt{intThreshold} and \texttt{removeNode} arguments. This helps to highlight variables with high VIVI scores, which is useful in settings with many predictors. The \texttt{intThreshold} argument filters edges with weight (i.e., VInt value) below a specified value and \texttt{removeNode} removes nodes with no connecting edges after thresholding interaction values. We can optionally cluster similar variables together with respect to their VIVI scores via the \texttt{cluster} argument, thereby aiding in the process of highlighting variables of interest. The \texttt{cluster} argument can take either a vector of cluster memberships for nodes or an appropriate \CRANpkg{igraph} clustering function.

We demonstrate network plots displaying VIVI values for the GBM fit. In Figure \ref{fig:networks1}, we show both a default network plot including all variables in (a) and a filtered and clustered network plot in (b). For the filtered plot we select VIVI values in the top decile. This selection allows us to focus only on the variables with the most impact on the response. The variables that remain are \(lstat\), \(nox\), \(rm\), \(crim\), \(dis\) (weighted mean of distances to five Boston employment centers), \(tax\) (full-value property-tax rate per \$10,000), and \(ptratio\) (pupil-teacher ratio by town). We then perform a hierarchical clustering treating variable interactions as similarities, with the goal of grouping together high-interaction variables. Here we manually select the number of groups we want to show via the \texttt{cutree} function (which cuts clustered data into a desired number of groups). Finally we rearrange the layout using \CRANpkg{igraph}. Here, \texttt{igraph::layout\_as\_star} places the first variable (deemed most relevant using the VIVI seriation process above) at the center, which in Figure \ref{fig:networks1} (b) emphasizes its key role as the most important predictor which also has the strongest interactions.

\begin{verbatim}
# default network plot for GBM
viviNetwork(viviGBst)

# clustered and filtered network for GBM
intVals <- viviGBst
diag(intVals) <- NA 

# select VIVI values in top 10%
impTresh <- quantile(diag(viviGBst),.9)
intThresh <- quantile(intVals,.9,na.rm=TRUE)
sv <- which(diag(viviGBst) > impTresh |
              apply(intVals, 1, max, na.rm=TRUE) > intThresh)
              
h <- hclust(-as.dist(viviGBst[sv,sv]), method = "single")

viviNetwork(viviGBst[sv,sv],
            intLims = c(0,1),
            impLims = c(0,8),
            cluster = cutree(h, k = 3), # specify number of groups
            layout = igraph::layout_as_star)
\end{verbatim}

In Figure \ref{fig:networks1}(a), when displaying all the variables, we can clearly identify which variables have the highest VIVI values. The large darker nodes of \(lstat\) and \(rm\) indicate their importance and the dark, thick connecting edge between \(lstat\) and \(nox\) tell us that these two variables strongly interact. In (b), after applying a hierarchical clustering, we can see the strongest mutual interactions have been grouped together for the GBM fit. Namely, \(lstat\), \(nox\), \(crim\), \(rm\), and \(dis\) are all grouped together. The remaining variables are individually clustered.

We provide a conversion of \texttt{vivid} matrix objects to a data frame via an \texttt{as.data.frame} method, as demonstrated below. This facilitates plotting with base R and \CRANpkg{ggplot2}, for example a barplot of either VImp or VInt values. Note that while \texttt{vivi} returns a matrix of class \texttt{vivid}, the class attribute was dropped when the matrix was re-ordered.

\begin{verbatim}
class(viviRf)<- c("vivid", class(viviRf)) 
head(as.data.frame(viviRf), 4)
\end{verbatim}

\begin{verbatim}
#>   Variable_1 Variable_2      Value Measure Row Col
#> 1      lstat      lstat 4.80970237    Vimp   1   1
#> 2        nox      lstat 0.06387693    Vint   2   1
#> 3         rm      lstat 0.15226649    Vint   3   1
#> 4       crim      lstat 0.44728494    Vint   4   1
\end{verbatim}

\hypertarget{GPDP}{%
\section{Partial Dependence and Individual Conditional Expectation Curves}\label{GPDP}}

\hypertarget{univariate-partial-dependence-plot}{%
\subsection{Univariate Partial Dependence Plot}\label{univariate-partial-dependence-plot}}

\begin{figure}

{\centering \includegraphics[width=1\linewidth]{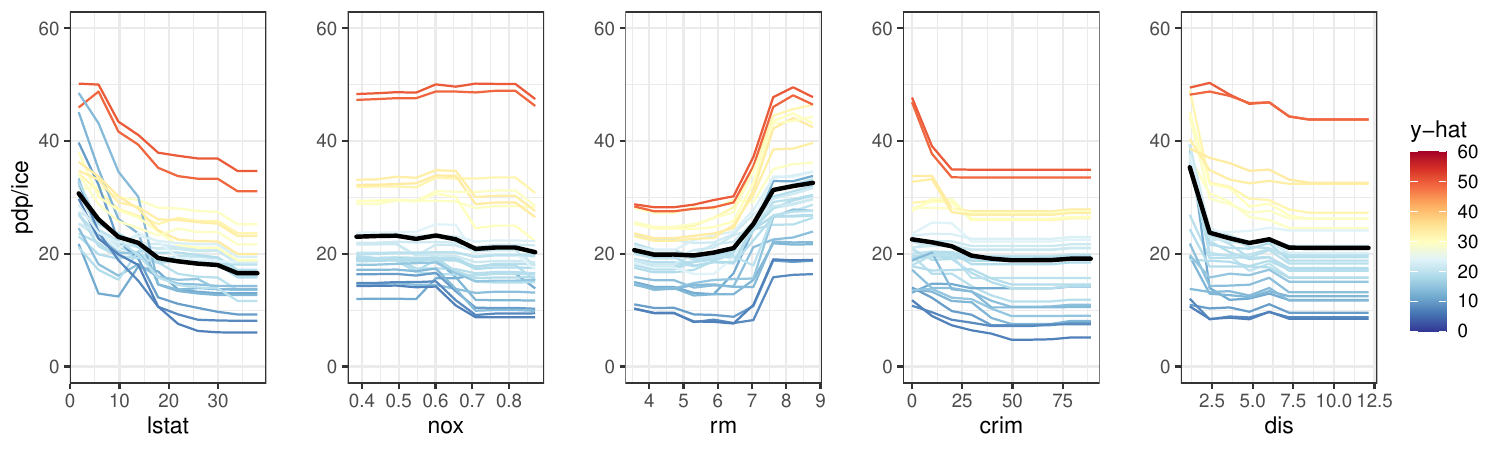} 

}

\caption{Partial dependence plots (black line) with individual conditional expectation curves (colored lines) of a GBM fit on the Boston housing data. The changing partial dependence and ICE curves of $lstat$ and $rm$ indicate that these variables have some impact on the response.}\label{fig:pdpRf}
\end{figure}

The \texttt{pdpVars} function constructs a grid of univariate PDPs with ICE curves for selected variables. We use ICE curves to assist in the identification of linear or non-linear effects. The fit, data frame used to train the model, and the name of the response variable are required inputs. In the code below, we show an example of the partial dependence and ICE curves for the first five features from the GBM \texttt{vivid} matrix, with output shown in Figure \ref{fig:pdpRf}. We use the custom GBM predict function given previously.

\begin{verbatim}
top5 <- colnames(viviGBst)[1:5]
pdpVars(data = Boston,
        fit = gbst,
        response = "medv",
        vars = top5,
        predictFun = pFun)
\end{verbatim}

All of our PDP variants handle categorical responses and predictors. The color palette is customized via the \texttt{pal} argument. In all of our PDPs, this defaults to a diverging palette which accentuates fitted values that differ from the average. Dark red and dark blue are used to indicate high and low values of \(\hat{y}\) respectively. The middle values are displayed in yellow. The \texttt{nIce} argument specifies the number of ICE curves to be drawn. This is either a single number specifying the number of observations to be sampled for the ICE curves, or a vector of row indices, an option that is useful for example to display ICE curves from particular classes. The default value for \texttt{nIce} is 30, which allows individual curves to be seen.

The ordering of the PDPs is taken from the ordering of variables in the data set, or may be specified via the \texttt{vars} argument. In Figure \ref{fig:pdpRf}, the ordering is taken directly from our seriated \texttt{vivid} matrix, thereby showing the top five most influential variables. As with the construction of the \texttt{vivid} matrix, the \texttt{gridSize} and \texttt{nmax} arguments determine the number of predictions.

In Figure \ref{fig:pdpRf} we can see from the changing PDP and ICE curves that \(lstat\) and \(rm\) have the clearest impact on the response, with the predicted median house price being higher for low values of \(lstat\) and high values of \(rm\). Additionally, the predicted median house price appears to be higher for low values of \(dis\) before leveling off at around 2.5. The remaining variables have generally flat partial dependence and ICE curves.

\hypertarget{generalized-pairs-partial-dependence-plot}{%
\subsection{Generalized Pairs Partial Dependence Plot}\label{generalized-pairs-partial-dependence-plot}}

The \texttt{pdpPairs} function creates a generalized pairs partial dependence plot (GPDP). In our GPDP, we use a matrix layout and plot the univariate partial dependence (with ICE curves) on the diagonal, bivariate partial dependence on the upper diagonal and a scatterplot of raw variable values on the lower diagonal, where all colors are assigned to points and ICE curves by the predicted \(\hat{y}\) value. In the case of categorical predictors, the partial dependence for each factor level is shown in the upper-diagonal (for an example of this, see Inglis, Parnell, and Hurley (2022a)). As with the univariate PDP, the fit, data frame used to train the model, and the name of the response variable are required inputs.

\begin{verbatim}
set.seed(1701)
pdpPairs(data = Boston,
         fit = gbst,
         response = "medv",
         gridSize = 20,
         nIce = 50,
         vars = top5,
         convexHull = TRUE,
         fitlims = "pdp",
         predictFun = pFun)
\end{verbatim}

\begin{figure}

{\centering \includegraphics[width=0.75\linewidth]{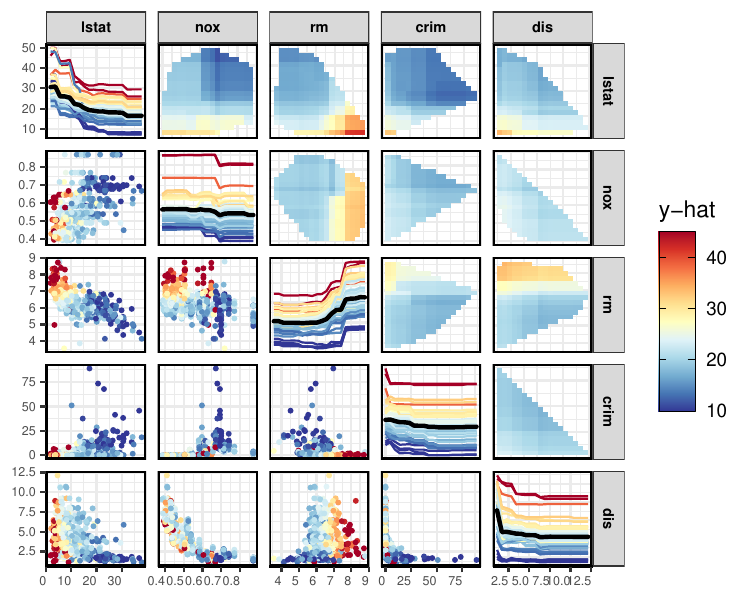} 

}

\caption{Filtered generalized pairs partial dependence plot for a GBM fit on the Boston housing data. From both the univariate and bivariate PDPs, we can see that $lstat$ and $rm$ have an impact on the response. As $lstat$ decreases and $rm$ increases, predicted median house price value goes up. The bivariate PDP of $lstat:nox$ shows that as $nox$ increases, the predicted value decreases.}\label{fig:gpdp}
\end{figure}

As with the univariate PDP, the ordering can be controlled via the \texttt{var} argument. By default, the ordering is taken from the order of the data. In the code above, we display only the interesting variables seen in previous plots by selecting the first five variables from our seriated \texttt{vivid} matrix. We also chose to display 50 ICE curves. As with \texttt{pdpVars}, additional arguments specify the color palette and number of ice curves, while \texttt{gridSize} and \texttt{nmax} determine the number of predictions.

For our GPDP, we follow the general design choices in \CRANpkg{vivid} and specify the range of predicted values to be mapped to the colors via the \texttt{fitlims} argument. We set the default fit range for the color map for the GPDP to the range of the collection of PDP surfaces with \texttt{fitlims\ =\ "pdp"}. The setting of this argument at its default value allows for maximum resolution of the bivariate PDPs. Since predictions for specific observations and ICE curves would likely exceed these bounds, the closest value within the color map's bounds is used to allocate colors. Alternatively \texttt{fitlims\ =\ \textquotesingle{}all"} specifies that limits are calculated as the full range of predictions shown.

In the upper diagonals we exclude extrapolated areas from the bivariate PDPs to prevent interpretation of the PDPs in areas where there is no data. The removal of extrapolated areas can be prevented by specifying \texttt{convexHull\ =\ FALSE}.

In Figure \ref{fig:gpdp}, in addition to the univariate PDPs, we capture the effects of the variables on the response via the bivariate PDP on the upper-diagonal and the distribution of the data in the lower-diagonal. The scatterplots are useful for determining if variables are highly correlated, as highly correlated variables may spuriously affect the partial dependence and give erroneous results (Apley and Zhu 2020). Of note are the variables \(lstat\) and \(rm\). We can clearly see that when the number of rooms (\(rm\)) is high and the percentage of lower status of the population (\(lstat\)) is low, the predicted \(\hat{y}\) median house price value is high. This is exemplified in the changing bivariate PDP.

\hypertarget{sec:ZPDP}{%
\section{Zen Partial Dependence Plots}\label{sec:ZPDP}}

The \texttt{pdpZen} function creates partial dependence plots utilizing a space-saving method based on graph Eulerians (Hierholzer and Wiener 1873). An Eulerian path, also known as an Eulerian trail, is a route that traverses each edge of a graph exactly once. When this path forms a closed loop, the traversal is referred to as an Eulerian tour. We call this display zen-partial dependence plots (ZPDP). The display is based on the zigzag expanded navigation plots, known as zenplots , available in the \CRANpkg{zenplots} package (Hofert and Oldford 2020). Zenplots were created to display paired graphs of high-dimensional data focusing on the most important 2D displays.
In our adaptation we show bivariate PDPs that focus on the variables with the largest interaction values in a compact zigzag layout, which is helpful when predictor space is high-dimensional.

The code below illustrates \texttt{pdpZen}, here displaying the first five variables from GBM's \texttt{vivid} matrix. Later we show an example focusing on high-interacting pairs of variables.
We use the same convention as our previous PDPs with regard to color palette and limits, grid size, and the number of rows considered for evaluation. The ZPDP also has a variable rug plot on each axis to avoid interpretation problems that may occur in the presence of skewness.

\begin{verbatim}
pdpZen(data = Boston,
       fit = gbst,
       response = "medv",
       convexHull = TRUE,
       zpath = top5,
       predictFun = pFun)
\end{verbatim}

\begin{figure}

{\centering \includegraphics[width=0.5\linewidth]{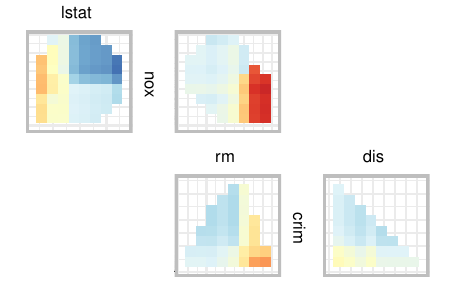} 

}

\caption{Zen partial dependence plot for the GBM fit on the Boston data. Here we display first five variables from the GBM's `vivid` matrix. Only plots for consecutive variables are shown.}\label{fig:zpdpGbm1}
\end{figure}

The argument \texttt{zpath} specifies the variables to be plotted, defaulting to all dataset variables aside from the response. In the code above, \texttt{zpath} is the vector \(lstat\), \(nox\), \(rm\), \(crim\) and \(dis\). The resulting plot shown in Figure \ref{fig:zpdpGbm1} presents the bivariate PDP for every consecutive pair of variables in a zigzag layout.

\hypertarget{sec:zp}{%
\subsection{\texorpdfstring{\texttt{Zen-paths}}{Zen-paths}}\label{sec:zp}}

ZPDP are most useful when the bivariate PDPs plotted are selected to be an interesting subset of all pairwise plots. To obtain this subset, we consider a network graph displaying VIVI values, such as that in Figure \ref{fig:networks1} (a). We then filter the edges below a selected interaction value, leaving only highly interacting variable pairs, as in Figure \ref{fig:networks1}(b). Our goal is to then build a ZPDP consisting of the bivariate plots represented by each edge of the thresholded graph. The \texttt{zPath} function creates a sequence or sequences of variable paths for use in \texttt{pdpZen}.

The \texttt{zPath} function takes four arguments. These are:
\texttt{viv} - a matrix of interaction values,
\texttt{cutoff} - exclude interaction values below this threshold,
\texttt{method} - a string indicating which method to use to create the path, and
\texttt{connect} - a logical value indicating if separate Eulerians should be connected

Two methods are provided, either \texttt{"greedy.weighted"} or \texttt{"strictly.weighted"}. The first option uses the greedy Eulerian path algorithm (C. B. Hurley and Oldford 2011, 2022) for connected graphs. This visits each edge at least once, beginning at the edge with the highest weight and traversing through the remaining edges, giving priority to the highest-weighted edge. Some edges may be visited more than once or additional edges may be visited if the number of nodes in the graph is not even. The second method \texttt{"strictly.weighted"} (provided by \CRANpkg{zenplot}) visits edges strictly in decreasing order by weight (here the interaction values). If \texttt{connect} is \texttt{TRUE} the sequences obtained by the strictly weighted method are concatenated to form a single path.

In the code below, we provide two examples of creating zen-paths, from the top 10\% of interaction scores in \texttt{viviGBst}.

\begin{verbatim}
intThresh <- quantile(intVals, .9, na.rm=TRUE)
# set zpaths with different parameters
zpGw  <- zPath(viv = viviGBst, cutoff = intThresh, method = "greedy.weighted")
zpGw
\end{verbatim}

\begin{verbatim}
#>  [1] "nox"     "lstat"   "dis"     "ptratio" "lstat"   "rm"      "crim"   
#>  [8] "lstat"   "tax"     "rm"      "nox"
\end{verbatim}

\begin{verbatim}
zpSw  <- zPath(viv = viviGBst, cutoff = intThresh, connect = FALSE, method = "strictly.weighted")
zpSw
\end{verbatim}

\begin{verbatim}
#> [[1]]
#> [1] "nox"   "lstat" "dis"  
#> 
#> [[2]]
#> [1] "lstat" "rm"    "nox"  
#> 
#> [[3]]
#> [1] "lstat" "crim"  "rm"   
#> 
#> [[4]]
#> [1] "ptratio" "lstat"   "tax"
\end{verbatim}

Our first created zen-path object, \texttt{zpGw}, uses the \texttt{greedy.weighted} method and visits each edge at exactly once. The second zen-path uses the \texttt{strictly.weighted} method with \texttt{connect\ =\ FALSE}.
\texttt{zpSw} consists of four unconnected paths.
The zenplots for two of these paths are constructed below.

\begin{verbatim}
pdpZen(data = Boston,
       fit = gbst,
       response = "medv",
       zpath = zpGw,
       convexHull = TRUE,
       predictFun = pFun) 

pdpZen(data = Boston,
       fit = gbst,
       response = "medv",
       zpath = zpSw,
       convexHull = TRUE,
       predictFun = pFun)
\end{verbatim}

\begin{figure}

{\centering \subfloat[\label{fig:zpdpGBM-1}]{\includegraphics[width=0.5\linewidth]{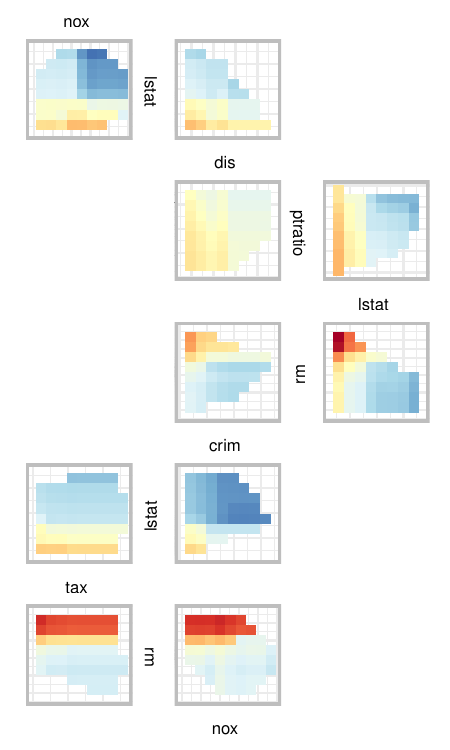} }\subfloat[\label{fig:zpdpGBM-2}]{\includegraphics[width=0.5\linewidth]{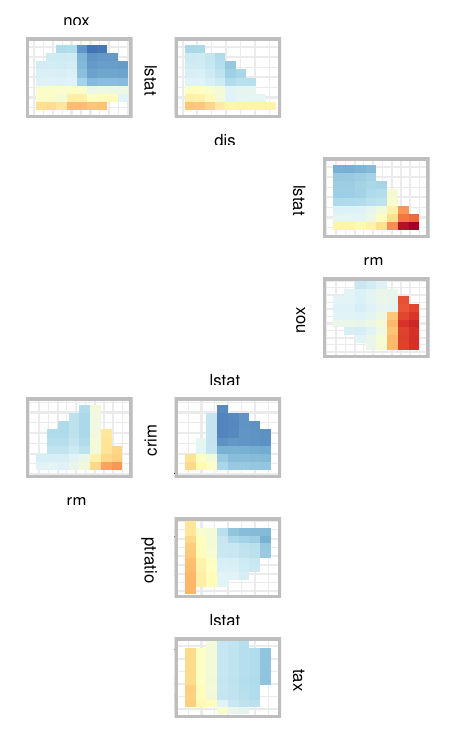} }

}

\caption{ZPDP for a GBM fit on the Boston data. In (a) the zpath is defined by the `greedy.weighted` sorting method. In (b), the sorting method is defined by the `strictly.weighted` method and is unconnected. For low values of $lstat$ and and high values of $rm$, predicted median house price value increases.}\label{fig:zpdpGBM}
\end{figure}

Note that there are 7 different variables involved in high interactions, which could be displayed in a \(7 \times 7\) GPDP, showing a total of 21 bivariate PDPs. But only 8 of these have VInt values above the 90\% quantile, and Figure \ref{fig:zpdpGBM}(b) using the \texttt{strictly.weighted} path shows just these bivariate PDPs compact layout. Using the \texttt{greedy.weighted} sorting method in (a) produces a smaller, neater plot but at the expense of including some plots that are not particularly interesting (for example the pair \(dis:ptratio\)). However, it should be noted that any of the arguments from the \texttt{zenplot} function from the \CRANpkg{zenplots} package can be used with the \texttt{pdpZen} function. These include multiple options for different or custom layouts.

\hypertarget{summary}{%
\section{Summary}\label{summary}}

We have presented a detailed exposition of our R package \CRANpkg{vivid} which contains a suite of integrated functions implementing algorithms and novel visualizations for exploring variable importance and variable interactions in machine learning models. Our techniques are intuitive, adaptable, easy to customize and facilitate model comparison. When building the \texttt{vivid} matrix to use in our heatmap and network visualizations, VIVI metrics that are model specific or model-agnostic may be employed. For measuring interactions we currently only provide the option to use the agnostic Friedman's \(H\)-statistic. However, as outlined in the \protect\hyperlink{data}{Calculating VIVI} section, the inclusion of different VIVI measures is easily possible.

Our \CRANpkg{vivid} package is a useful addition to the other packages in the area of model visualization, such as those discussed in the \protect\hyperlink{introduction}{Introduction} section. Our heatmap and network plots efficiently determine which variables have the greatest impact on the response. When coupled with the seriation, filtering, and clustering techniques, these visualizations enhance the interpretation of ML predictions. Our GPDP and ZPDP can be used to provide a thorough examination of the behavior of a fitted ML model by examining the individual variable effects and their pairwise interactions. These plots combine the bivariate PDP, ICE curves, and scatterplots of the raw variable values. They further allow focusing on subsets of variables with high VInt, and so allow us to efficiently explore a fitted ML model by focusing attention to only the most crucial aspects.

It has been noted that the presence of correlated variables can lead to biased VIVI measures. For example, Friedman and Popescu (2008) note that when subsets of variables are highly correlated, it becomes difficult to distinguish between low and higher order interactions among them in a straightforward manner when using the \(H\)-statistic. Similarly, Fisher, Rudin, and Dominici (2019) state that for permutation variable importance, the presence of highly correlated variables can affect the measured variable importance and propose using a method of conditional importance, such as that implemented for conditional random forests in the \CRANpkg{party} package. Our GPDP can be used to identify any evident correlations by observing the plot of the data and the PDP with convex hull. However, careful consideration should be employed when interpreting the VIVI measures from \CRANpkg{vivid} and we recommend evaluating any potential correlation between variables in conjunction with our proposed visualizations. Several \texttt{R} packages are available for assessing and visualising correlations, such as \CRANpkg{corrplot} (Wei and Simko 2021) and \CRANpkg{corrgrapher} (Morgen and Biecek 2020).

For future work, the inclusion of other model summaries could be incorporated into \CRANpkg{vivid}, such as the interaction statistics described in Greenwell, Boehmke, and McCarthy (2018) or the use of accumulated local effects (ALE) of Apley and Zhu (2020). This latter method was created to address bias problems with partial dependency functions and could be used in place of the bivariate PDPs seen in both the GPDP and ZPDP. However the calculation of an agnostic, easily interpretable variable interaction measure that accounts for correlated variables remains an ongoing research goal. Allowing a user to view the variation across replicates when performing permutation importance could also be incorporated. This could be added into the heatmap graphic by way of a value suppressing uncertainty palette (Correll, Moritz, and Heer 2018), where the uncertainty in included in the visualization (for example, see Inglis, Parnell, and Hurley (2022b)). Additionally, providing alternative metrics such as AUC or accuracy for computation of permutation importance in classification, as well as computations and visualizations of higher order interactions, interactive graphics and incorporating multivariate response variables are interesting areas for future research.

\section{Funding:} 
Alan Inglis and Andrew Parnell’s work was supported by a Science Foundation Ireland Career Development Award grant 17/CDA/4695. In addition Andrew Parnell’s work was supported by: an investigator award
(16/IA/4520); a Marine Research Programme funded by the Irish Government, co-financed by the European Regional Development Fund (Grant-
Aid Agreement No. PBA/CC/18/01); European Union’s Horizon 2020
research and innovation programme InnoVar under grant agreement No
818144; SFI Centre for Research Training in Foundations of Data Science 18CRT/6049, and SFI Research Centre awards I-Form 16/RC/3872 and Insight 12/RC/2289\_P2. For the purpose of Open Access, the author
has applied a CC BY public copyright licence to any Author Accepted
Manuscript version arising from this submission.

\hypertarget{references}{%
\section*{References}\label{references}}
\addcontentsline{toc}{section}{References}

\hypertarget{refs}{}
\begin{CSLReferences}{1}{0}
\leavevmode\vadjust pre{\hypertarget{ref-antunes2018fairness}{}}%
Antunes, Nuno, Leandro Balby, Flavio Figueiredo, Nuno Lourenco, Wagner Meira, and Walter Santos. 2018. {``Fairness and Transparency of Machine Learning for Trustworthy Cloud Services.''} In \emph{2018 48th Annual IEEE/IFIP International Conference on Dependable Systems and Networks Workshops (DSN-w)}, 188--93. IEEE.

\leavevmode\vadjust pre{\hypertarget{ref-apley2020visualizing}{}}%
Apley, Daniel W, and Jingyu Zhu. 2020. {``Visualizing the Effects of Predictor Variables in Black Box Supervised Learning Models.''} \emph{Journal of the Royal Statistical Society: Series B (Statistical Methodology)} 82 (4): 1059--86.

\leavevmode\vadjust pre{\hypertarget{ref-dalex}{}}%
Biecek, Przemyslaw. 2018. {``DALEX: Explainers for Complex Predictive Models in {R}.''} \emph{Journal of Machine Learning Research} 19 (84): 1--5. \url{https://jmlr.org/papers/v19/18-416.html}.

\leavevmode\vadjust pre{\hypertarget{ref-ingred}{}}%
Biecek, Przemyslaw, and Hubert Baniecki. 2023. \emph{Ingredients: Effects and Importances of Model Ingredients}. \url{https://CRAN.R-project.org/package=ingredients}.

\leavevmode\vadjust pre{\hypertarget{ref-biecek2021explanatory}{}}%
Biecek, Przemyslaw, and Tomasz Burzykowski. 2021. \emph{Explanatory Model Analysis: Explore, Explain and Examine Predictive Models}. Chapman; Hall/CRC.

\leavevmode\vadjust pre{\hypertarget{ref-mlr}{}}%
Bischl, Bernd, Michel Lang, Lars Kotthoff, Julia Schiffner, Jakob Richter, Erich Studerus, Giuseppe Casalicchio, and Zachary M. Jones. 2016. {``{mlr}: Machine Learning in {R}.''} \emph{Journal of Machine Learning Research} 17 (170): 1--5. \url{https://jmlr.org/papers/v17/15-066.html}.

\leavevmode\vadjust pre{\hypertarget{ref-breiman2001randomforest}{}}%
Breiman, Leo. 2001. {``Random Forests.''} \emph{Machine Learning} 45 (1): 5--32. \url{https://doi.org/10.1023/A:1010933404324}.

\leavevmode\vadjust pre{\hypertarget{ref-chen2012random}{}}%
Chen, Xi, and Hemant Ishwaran. 2012. {``Random Forests for Genomic Data Analysis.''} \emph{Genomics} 99 (6): 323--29.

\leavevmode\vadjust pre{\hypertarget{ref-chipman2010bart}{}}%
Chipman, Hugh A, Edward I George, and Robert E McCulloch. 2010. {``BART: Bayesian Additive Regression Trees.''} \emph{The Annals of Applied Statistics} 4 (1): 266--98.

\leavevmode\vadjust pre{\hypertarget{ref-vsup}{}}%
Correll, Michael, Dominik Moritz, and Jeffrey Heer. 2018. {``Value-{S}uppressing {U}ncertainty {P}alettes.''} In \emph{Proceedings of the 2018 CHI Conference on Human Factors in Computing Systems}, 1--11.

\leavevmode\vadjust pre{\hypertarget{ref-igraph}{}}%
Csardi, Gabor, and Tamas Nepusz. 2006. {``The Igraph Software Package for Complex Network Research.''} \emph{InterJournal} Complex Systems: 1695. \url{https://igraph.org}.

\leavevmode\vadjust pre{\hypertarget{ref-dendserPaper}{}}%
Earle, Denise, and Catherine Hurley. 2015. {``Advances in Dendrogram Seriation for Application to Visualization.''} \emph{Journal of Computational and Graphical Statistics} 24 (March). \url{https://doi.org/10.1080/10618600.2013.874295}.

\leavevmode\vadjust pre{\hypertarget{ref-felzmann2019transparency}{}}%
Felzmann, Heike, Eduard Fosch Villaronga, Christoph Lutz, and Aurelia Tamò-Larrieux. 2019. {``Transparency You Can Trust: Transparency Requirements for Artificial Intelligence Between Legal Norms and Contextual Concerns.''} \emph{Big Data \& Society} 6 (1): 2053951719860542.

\leavevmode\vadjust pre{\hypertarget{ref-fisher2019all}{}}%
Fisher, Aaron, Cynthia Rudin, and Francesca Dominici. 2019. {``All Models Are Wrong, but Many Are Useful: Learning a Variable's Importance by Studying an Entire Class of Prediction Models Simultaneously.''} \emph{J. Mach. Learn. Res.} 20 (177): 1--81.

\leavevmode\vadjust pre{\hypertarget{ref-netviz1}{}}%
Freeman, Linton C. 2000. {``Visualizing Social Networks.''} \emph{Journal of Social Structure} 1 (1): 4.

\leavevmode\vadjust pre{\hypertarget{ref-Friedpdp}{}}%
Friedman, J. H. 2000. {``Greedy Function Approximation: A Gradient Boosting Machine.''} \emph{The Annals of Statistics} 29 (November). \url{https://doi.org/10.1214/aos/1013203451}.

\leavevmode\vadjust pre{\hypertarget{ref-Friedmans_H}{}}%
Friedman, J. H., and B. E. Popescu. 2008. {``Predictive Learning via Rule Ensembles.''} \emph{The Annals of Applied Statistics.}, 916--54.

\leavevmode\vadjust pre{\hypertarget{ref-iceR}{}}%
Goldstein, Alex, Adam Kapelner, Justin Bleich, and Emil Pitkin. 2015. {``Peeking Inside the Black Box: Visualizing Statistical Learning with Plots of Individual Conditional Expectation.''} \emph{Journal of Computational and Graphical Statistics} 24 (1): 44--65. \url{https://doi.org/10.1080/10618600.2014.907095}.

\leavevmode\vadjust pre{\hypertarget{ref-grange2018random}{}}%
Grange, Stuart K, David C Carslaw, Alastair C Lewis, Eirini Boleti, and Christoph Hueglin. 2018. {``Random Forest Meteorological Normalisation Models for Swiss PM 10 Trend Analysis.''} \emph{Atmospheric Chemistry and Physics} 18 (9): 6223--39.

\leavevmode\vadjust pre{\hypertarget{ref-pdpR}{}}%
Greenwell, Brandon M. 2017. {``{pdp}: An r Package for Constructing Partial Dependence Plots.''} \emph{The R Journal} 9 (1): 421--36. \url{https://journal.r-project.org/archive/2017/RJ-2017-016/index.html}.

\leavevmode\vadjust pre{\hypertarget{ref-vip}{}}%
Greenwell, Brandon M., and Bradley C. Boehmke. 2020. {``Variable Importance Plots---an Introduction to the {vip} Package.''} \emph{The R Journal} 12 (1): 343--66. \url{https://doi.org/10.32614/RJ-2020-013}.

\leavevmode\vadjust pre{\hypertarget{ref-greenwell2018simple}{}}%
Greenwell, Brandon M., Bradley C. Boehmke, and Andrew J. McCarthy. 2018. {``A Simple and Effective Model-Based Variable Importance Measure.''} \emph{arXiv Preprint arXiv:1805.04755}.

\leavevmode\vadjust pre{\hypertarget{ref-boston}{}}%
Harrison Jr, David, and Daniel L Rubinfeld. 1978. {``Hedonic Housing Prices and the Demand for Clean Air.''} \emph{Journal of Environmental Economics and Management} 5 (1): 81--102.

\leavevmode\vadjust pre{\hypertarget{ref-elementsStats}{}}%
Hastie, T., R. Tibshirani, and J. Friedman. 2009. \emph{The Elements of Statistical Learning: Data Mining, Inference, and Prediction.} 2nd ed. Springer Series in Statistics. Springer-Verlag.

\leavevmode\vadjust pre{\hypertarget{ref-netviz2}{}}%
Herman, Ivan, Guy Melançon, and M Scott Marshall. 2000. {``Graph Visualization and Navigation in Information Visualization: A Survey.''} \emph{IEEE Transactions on Visualization and Computer Graphics} 6 (1): 24--43.

\leavevmode\vadjust pre{\hypertarget{ref-hierholzer1873moglichkeit}{}}%
Hierholzer, Carl, and Chr Wiener. 1873. {``{Ü}ber Die m{ö}glichkeit, Einen Linienzug Ohne Wiederholung Und Ohne Unterbrechung Zu Umfahren.''} \emph{Mathematische Annalen} 6 (1): 30--32.

\leavevmode\vadjust pre{\hypertarget{ref-zenplot}{}}%
Hofert, Marius, and Wayne Oldford. 2020. {``Zigzag Expanded Navigation Plots in {R}: The {R} Package {zenplots}.''} \emph{Journal of Statistical Software} 95 (4): 1--44. \url{https://doi.org/10.18637/jss.v095.i04}.

\leavevmode\vadjust pre{\hypertarget{ref-PairViz1}{}}%
Hurley, Catherine B., and R. W. Oldford. 2011. {``Eulerian Tour Algorithms for Data Visualization and the PairViz Package.''} \emph{Computational Statistics} 26 (December): 613--33. \url{https://doi.org/10.1007/s00180-011-0229-5}.

\leavevmode\vadjust pre{\hypertarget{ref-pviz}{}}%
---------. 2022. \emph{PairViz: Visualization Using Graph Traversal}. \url{https://CRAN.R-project.org/package=PairViz}.

\leavevmode\vadjust pre{\hypertarget{ref-condvis2}{}}%
Hurley, Catherine, Mark OConnell, and Katarina Domijan. 2022. \emph{Condvis2: Interactive Conditional Visualization for Supervised and Unsupervised Models in Shiny}.

\leavevmode\vadjust pre{\hypertarget{ref-lime}{}}%
Hvitfeldt, Emil, Thomas Lin Pedersen, and Michaël Benesty. 2022. \emph{Lime: Local Interpretable Model-Agnostic Explanations}. \url{https://CRAN.R-project.org/package=lime}.

\leavevmode\vadjust pre{\hypertarget{ref-inglis2022visualizing}{}}%
Inglis, Alan, Andrew Parnell, and Catherine B Hurley. 2022a. {``Visualizing Variable Importance and Variable Interaction Effects in Machine Learning Models.''} \emph{Journal of Computational and Graphical Statistics} 31 (3): 766--78. \url{https://doi.org/10.1080/10618600.2021.2007935}.

\leavevmode\vadjust pre{\hypertarget{ref-bartvis}{}}%
Inglis, Alan, Andrew Parnell, and Cathrine Hurley. 2022b. {``Visualizations for Bayesian Additive Regression Trees.''} \emph{arXiv Preprint arXiv:2208.08966}.

\leavevmode\vadjust pre{\hypertarget{ref-ishwaran2010high}{}}%
Ishwaran, Hemant, Udaya B Kogalur, Eiran Z Gorodeski, Andy J Minn, and Michael S Lauer. 2010. {``High-Dimensional Variable Selection for Survival Data.''} \emph{Journal of the American Statistical Association} 105 (489): 205--17.

\leavevmode\vadjust pre{\hypertarget{ref-parsnip}{}}%
Kuhn, Max, and Davis Vaughan. 2022. \emph{Parsnip: A Common API to Modeling and Analysis Functions}. \url{https://CRAN.R-project.org/package=parsnip}.

\leavevmode\vadjust pre{\hypertarget{ref-mlr3}{}}%
Lang, Michel, Martin Binder, Jakob Richter, Patrick Schratz, Florian Pfisterer, Stefan Coors, Quay Au, Giuseppe Casalicchio, Lars Kotthoff, and Bernd Bischl. 2019. {``{mlr3}: A Modern Object-Oriented Machine Learning Framework in {R}.''} \emph{Journal of Open Source Software}, December. \url{https://doi.org/10.21105/joss.01903}.

\leavevmode\vadjust pre{\hypertarget{ref-EIX}{}}%
Maksymiuk, Szymon, Ewelina Karbowiak, and Przemyslaw Biecek. 2021. \emph{EIX: Explain Interactions in 'XGBoost'}. \url{https://CRAN.R-project.org/package=EIX}.

\leavevmode\vadjust pre{\hypertarget{ref-svmPack}{}}%
Meyer, David, Evgenia Dimitriadou, Kurt Hornik, Andreas Weingessel, and Friedrich Leisch. 2021. \emph{E1071: Misc Functions of the Department of Statistics, Probability Theory Group (Formerly: E1071), TU Wien}. \url{https://CRAN.R-project.org/package=e1071}.

\leavevmode\vadjust pre{\hypertarget{ref-molnar2022book}{}}%
Molnar, Christoph. 2022. \emph{Interpretable Machine Learning: A Guide for Making Black Box Models Explainable}. 2nd ed. \url{https://christophm.github.io/interpretable-ml-book}.

\leavevmode\vadjust pre{\hypertarget{ref-iml}{}}%
Molnar, Christoph, Bernd Bischl, and Giuseppe Casalicchio. 2018. {``Iml: An r Package for Interpretable Machine Learning.''} \emph{JOSS} 3 (26): 786. \url{https://doi.org/10.21105/joss.00786}.

\leavevmode\vadjust pre{\hypertarget{ref-corrgrapher}{}}%
Morgen, Pawel, and Przemyslaw Biecek. 2020. \emph{Corrgrapher: Explore Correlations Between Variables in a Machine Learning Model}. \url{https://CRAN.R-project.org/package=corrgrapher}.

\leavevmode\vadjust pre{\hypertarget{ref-murray2009methods}{}}%
Murray, Kim, and Mary M Conner. 2009. {``Methods to Quantify Variable Importance: Implications for the Analysis of Noisy Ecological Data.''} \emph{Ecology} 90 (2): 348--55.

\leavevmode\vadjust pre{\hypertarget{ref-RFE}{}}%
Paluszynska, Aleksandra, Przemyslaw Biecek, and Yue Jiang. 2020. \emph{randomForestExplainer: Explaining and Visualizing Random Forests in Terms of Variable Importance}. \url{https://CRAN.R-project.org/package=randomForestExplainer}.

\leavevmode\vadjust pre{\hypertarget{ref-varImpPack}{}}%
Probst, Philipp. 2020. \emph{varImp: {RF} {V}ariable {I}mportance for {A}rbitrary {M}easures}. \url{https://CRAN.R-project.org/package=varImp}.

\leavevmode\vadjust pre{\hypertarget{ref-netviz3}{}}%
Purchase, Helen. 1997. {``Which Aesthetic Has the Greatest Effect on Human Understanding?''} In \emph{Graph Drawing}, 97:248--61.

\leavevmode\vadjust pre{\hypertarget{ref-limeOG}{}}%
Ribeiro, Marco Tulio, Sameer Singh, and Carlos Guestrin. 2016. {``" Why Should i Trust You?" Explaining the Predictions of Any Classifier.''} In \emph{Proceedings of the 22nd ACM SIGKDD International Conference on Knowledge Discovery and Data Mining}, 1135--44.

\leavevmode\vadjust pre{\hypertarget{ref-kknn}{}}%
Schliep, Klaus, and Klaus Hechenbichler. 2016. \emph{Kknn: Weighted k-Nearest Neighbors}. \url{https://CRAN.R-project.org/package=kknn}.

\leavevmode\vadjust pre{\hypertarget{ref-party}{}}%
Strobl, Carolin, Anne-Laure Boulesteix, Thomas Kneib, Thomas Augustin, and Achim Zeileis. 2008. {``Conditional Variable Importance for Random Forests.''} \emph{BMC Bioinformatics} 9 (307). \url{https://doi.org/10.1186/1471-2105-9-307}.

\leavevmode\vadjust pre{\hypertarget{ref-sugiyama1981methods}{}}%
Sugiyama, Kozo, Shojiro Tagawa, and Mitsuhiko Toda. 1981. {``Methods for Visual Understanding of Hierarchical System Structures.''} \emph{IEEE Transactions on Systems, Man, and Cybernetics} 11 (2): 109--25.

\leavevmode\vadjust pre{\hypertarget{ref-mass}{}}%
Venables, W. N., and B. D. Ripley. 2002. \emph{Modern Applied Statistics with s}. Fourth. New York: Springer. \url{https://www.stats.ox.ac.uk/pub/MASS4/}.

\leavevmode\vadjust pre{\hypertarget{ref-corrplot2021}{}}%
Wei, Taiyun, and Viliam Simko. 2021. \emph{R Package 'Corrplot': Visualization of a Correlation Matrix}. \url{https://github.com/taiyun/corrplot}.

\leavevmode\vadjust pre{\hypertarget{ref-ggplot2}{}}%
Wickham, Hadley. 2016. \emph{Ggplot2: Elegant Graphics for Data Analysis}. Springer-Verlag New York. \url{https://ggplot2.tidyverse.org}.

\leavevmode\vadjust pre{\hypertarget{ref-ranger}{}}%
Wright, Marvin N., and Andreas Ziegler. 2017. {``{ranger}: A Fast Implementation of Random Forests for High Dimensional Data in {C++} and {R}.''} \emph{Journal of Statistical Software} 77 (1): 1--17. \url{https://doi.org/10.18637/jss.v077.i01}.

\leavevmode\vadjust pre{\hypertarget{ref-colorspaceJSS}{}}%
Zeileis, Achim, Jason C. Fisher, Kurt Hornik, Ross Ihaka, Claire D. McWhite, Paul Murrell, Reto Stauffer, and Claus O. Wilke. 2020. {``Colorspace: A Toolbox for Manipulating and Assessing Colors and Palettes.''} \emph{Journal of Statistical Software, Articles} 96 (1): 1--49. \url{https://doi.org/10.18637/jss.v096.i01}.

\end{CSLReferences}

\bibliography{vivid.bib}

\address{%
Alan Inglis\\
Maynooth University\\%
Department of Mathematics and Statistics\\ Maynooth, Ireland\\
\textit{ORCiD: \href{https://orcid.org/0000-0002-1151-6657}{0000-0002-1151-6657}}\\%
\href{mailto:alan.n.inglis@gmail.com}{\nolinkurl{alan.n.inglis@gmail.com}}%
}

\address{%
Andrew Parnell\\
Maynooth University\\%
Hamilton Institute\\ Maynooth, Ireland\\
\textit{ORCiD: \href{https://orcid.org/0000-0001-7956-7939}{0000-0001-7956-7939}}\\%
\href{mailto:andrew.parnell@mu.ie}{\nolinkurl{andrew.parnell@mu.ie}}%
}

\address{%
Catherine Hurley\\
Maynooth University\\%
Department of Mathematics and Statistics\\ Maynooth, Ireland\\
\textit{ORCiD: \href{https://orcid.org/0000-0003-2758-5531}{0000-0003-2758-5531}}\\%
\href{mailto:catherine.hurley@mu.ie}{\nolinkurl{catherine.hurley@mu.ie}}%
}

\end{article}

\end{document}